# Light propagation through metamaterial temporal slabs: reflection, refraction, and special cases

**DAVIDE RAMACCIA*, ALESSANDRO TOSCANO, FILIBERTO BILOTTI**

[1]*Department of Engineering, ROMA TRE University, Rome, 00146-Italy*
**Corresponding author: davide.ramaccia@uniroma3.it*

**Time-varying metamaterials are artificial materials whose electromagnetic properties change over time. Similar to a spatial medium discontinuity, a sudden change in time of the metamaterial refractive index induces the generation of a reflected and refracted light waves. The relationship between the incident and emerging fields at one temporal interface has been subject of investigation in earlier studies. Here, we extend the study to a temporal slab, i.e., a uniform homogeneous medium that is present in the whole space for a limited time. The scattering coefficients have been derived as a function of the refractive indices and application time, demonstrating that the response of the temporal slab can be controlled through the application time, which acts similarly to the electrical thickness of conventional spatial slabs. The results reported in this letter pave the way to novel devices based on temporal discontinuities, such as temporal matching networks, Bragg grating, dielectric mirrors, which exhibit zero space occupancy by exploiting the time dimension, instead of the spatial one.**

The reflection and refraction of light, or electromagnetic radiation in general, at an interface between two different optical media is one of the basic and well-known concepts in physics and electromagnetic field theory [1]. The relationship between the amplitude of the reflected and refracted electromagnetic waves with respect to the incident one is described through the Fresnel equations, which represent one of the building blocks of modern optics. During the last century, such equations have been continuously updated and generalized, taking into account the presence of moving interfaces [2], negative refractive indices [3,4], and weak spatial dispersion [5], just to name a few. It is worth highlighting that the aforementioned cases are all related to the more familiar concept of spatial discontinuity, *i.e.* two semi-infinite mediums that shares a common surface at which reflection and refraction occur. However, as originally speculated by F. Morgenthaler in 1958, a medium discontinuity may occur also in the temporal domain [6].

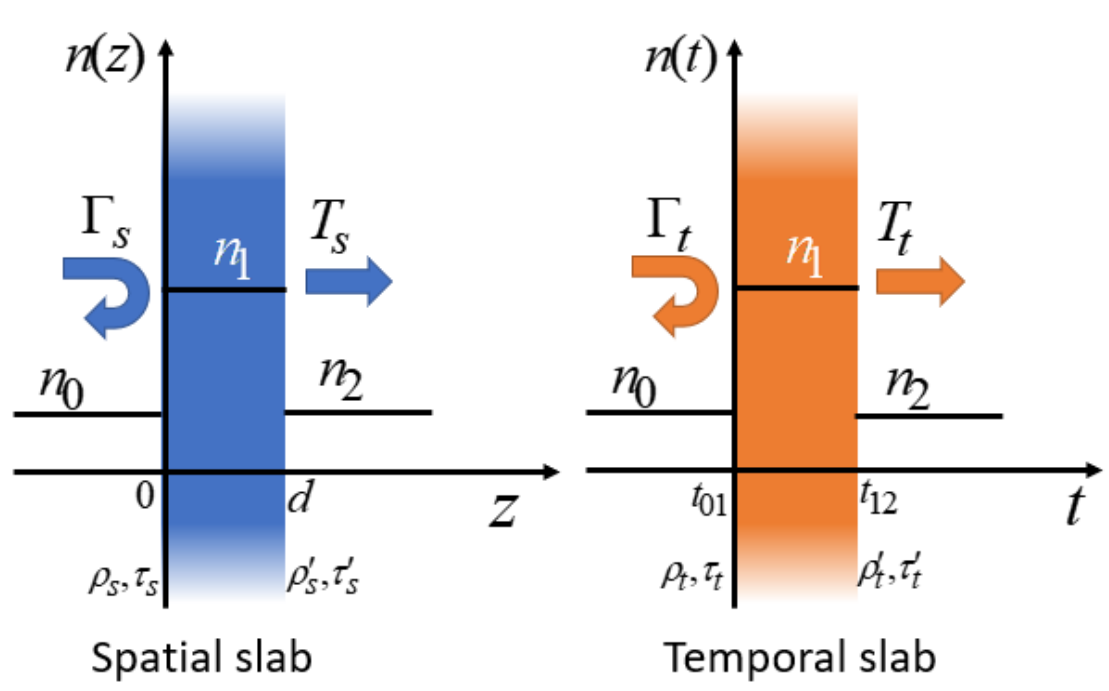

Fig. 1. Schematic illustration of a spatial (left) and temporal (right) material slab. The spatial slab consists of a layer of material with refractive index $n_1$ between two semi-infinite media with refractive indices $n_0$ and $n_2$; the temporal slab is a change of the material in all the space for a limited time interval $\Delta t$ .

In this case, the entire medium where light is propagating suddenly changes its refractive index, creating *de facto* a temporal interface between two different materials. F. Morgenthaler in [6] and Y. Xiao in [7] derived the relationship between the "reflected" and "refracted" waves with respect to the "incident" wave, i.e. the waves after and before the temporal discontinuity, respectively, formulating the corresponding temporal version of the Fresnel equations. What is interesting in these equations is that they preserve somehow the expected relation between the waves before and after the interface, but also some deep and fundamental differences can be identified. First, it was predicted that the temporal frequency of the incident wave shifts linearly with changes in the refractive index, but the spatial frequency, *i.e.* the wavevector, is conserved. This is totally dual to what happens at a spatial interface, where, on the contrary, the temporal frequency is always conserved, but the momentum of the refracted wave is scaled with the changes in the refractive index. Second, in temporal discontinuities, not only the refracted, but also the reflected wave exhibits a shift of the temporal frequency, since both waves propagate in the same medium after the discontinuity.

Although several interesting wave phenomena can be envisioned at the temporal interfaces, they were only marginally investigated in the past [8–11]. Only recently, the advent of space-time and time-varying metamaterials [12] has stimulated the interest in the anomalous interaction between the waves and time-varying metamaterial, enabling the implementation of some novel interesting components, such as non-reciprocal Bragg-gratings [13,14] and antennas [15–17], optical isolators [18], and Doppler cloaks [19]. Moreover, Engheta et al. in [20,21] investigated on the propagation in a temporal metamaterial whose refractive index suddenly changes to negative values, and the possibility to achieve temporal coatings.

In this letter, we propose the systematic study of the scattering response from a double-interface time-varying material, *i.e.*, a uniform homogeneous medium that is present in the whole space for a limited amount of time. This can be considered as the temporal version of a dielectric slab between two semi-infinite media, as shown in Fig. 1. Therefore, we refer to it as a *temporal slab*, which helps maintaining the connection to the corresponding spatial version. The aim of this letter, thus, is to analytically derive the total reflection and transmission coefficients and identify the special cases for achieving anomalous reflection and transmission through a metamaterial temporal slab. The analysis is conducted under the assumption of the harmonic time dependence $\exp[i\omega t]$.

*Theoretical analysis* - Let us consider a plane wave propagating in an infinite non-magnetic medium with refractive index $n_0 = \sqrt{\varepsilon_0}$. Its propagation constant is $k_0 = n_0\omega_0/c$. Since the temporal medium discontinuities conserve the wave momentum, but not the angular frequency, in the following the propagation term $\exp[i\underline{k}_0 \cdot \underline{r}]$ is suppressed. The instantaneous amplitude of the "incident" electric field is $E_i = E_0 e^{i\omega_0 t}$. At the instant of time $t = t_{01}$, the refractive index of the whole medium changes suddenly to $n_1 = \sqrt{\varepsilon_1}$, generating a transmitted wave $E_{\tau 1}$ that propagates in forward time-direction and a reflected wave $E_{\rho 1}$ that propagates in backward time-direction:

$$E_{\tau 1} = \tau_t E_0 e^{i\omega_1(t-t_{01})}$$
$$E_{\rho 1} = \rho_t E_0 e^{-i\omega_1(t-t_{01})} \qquad (1)$$

In (1), the required, and unfeasible, negative direction of the temporal axis for the reflected waves corresponds to a backward-propagating wave, *i.e.*, propagating in the opposite direction with respect to the "incident" one. Moreover, similar to the change of the wavevector after a spatial discontinuity, the angular frequency changes, as well, from $\omega_0$ to $\omega_1$. They are related to each other through the multiplying scaling factor $\xi_{01} = n_0/n_1$, that is nothing but the *time-dilation factor* related to the change in the speed of the wave after the first temporal boundary [7]. Finally, the amplitude of the two waves in (1) are dependent on the temporal Fresnel coefficients [6,7,12] at this first temporal interface:

$$\tau_t = \frac{1}{2}\frac{n_0}{n_1}\frac{Z_1 + Z_0}{Z_0}; \qquad \rho_t = \frac{1}{2}\frac{n_0}{n_1}\frac{Z_1 - Z_0}{Z_0}, \qquad (2)$$

where $Z_{0,1} = Z/n_{0,1}$ and $Z \simeq 377\Omega$ is the free-space impedance. At the time $t = t_{12}$, the refractive index of the medium again changes suddenly. The whole space exhibits now the refractive index $n_2 = \sqrt{\varepsilon_2}$. The two waves $E_{\tau 1}, E_{\rho 1}$ split again and each of them generates, in turn, a forward and backward wave, both at frequency $\omega_2 = \xi_{12}\omega_1$, with $\xi_{12} = n_1/n_2$, whose amplitudes are again related to the temporal Fresnel coefficients $\rho_t', \tau_t'$:

$$E_{\tau 2} = \left(\tau_t' E_{\tau 1}\Big|_{t=t_{12}} + \rho_t' E_{\rho 1}\Big|_{t=t_{12}}\right) e^{i\omega_2(t-t_{12})}$$
$$E_{r2} = \left(\tau_t' E_{\rho 1}\Big|_{t=t_{12}} + \rho_t' E_{\tau 1}\Big|_{t=t_{12}}\right) e^{-i\omega_2(t-t_{12})} \qquad (3)$$

where

$$\tau_t' = \frac{1}{2}\frac{n_1}{n_2}\frac{Z_2 + Z_1}{Z_1}, \quad \rho_t' = \frac{1}{2}\frac{n_1}{n2}\frac{Z_2 - Z_1}{Z_1}. \qquad (4)$$

Therefore, it is clear from (3) that a total of four scattering contributions are now propagating in the medium $n_2$. Substituting (1) into (3), we can derive the total temporal reflection and transmission coefficients as:

$$\Gamma_t = e^{i\xi\omega_0\Delta t}\left(\tau_t\rho_t' + \tau_t'\rho_t e^{-i2\xi\omega_0\Delta t}\right), \qquad (5)$$

$$T_t = e^{i\xi\omega_0\Delta t}\left(\tau_t\tau_t' + \rho_t'\rho_t e^{-i2\xi\omega_0\Delta t}\right), \qquad (6)$$

where $\Delta t = t_{12} - t_{01}$, and $\xi = \xi_{01}$, *i.e.* the temporal extension of the slab and the time-dilation factor caused by the first interface, respectively. Exploiting the analogy with a spatial slab, we can notice that the temporal slab exhibits a total reflection and transmission coefficients that resemble the ones for a spatial slab, where, however, the electrical thickness $d/\lambda$ is replaced by its effective temporal thickness $\xi\Delta t/T$ with $T = 2\pi/\omega_0$. The total scattering coefficients in (5)-(6) give us the possibility to design the temporal slab for achieving special transmission and reflection cases. In the following, we consider the special cases of zero reflection ($\Gamma_t = 0$ ) and zero transmission ($T_t = 0$ ) temporal slabs.

*Zero reflection/transmission condition* – The zeros of (5) correspond to a reflection-less interface at the incident operative frequency $\omega_0$. It is evident from (5) that a zero will occur if the bracketed term vanishes, which gives the condition:

$$e^{-i2\xi\omega_0\Delta t} = -\frac{\tau_t\rho_t'}{\tau_t'\rho_t} \qquad (7)$$

Because the right-hand side is real-valued and the left-hand side has unit magnitude, this condition can be satisfied only in the following two cases:

$$e^{-i2\xi\omega_0\Delta t} = +1 \;\Rightarrow\; \tau_t\rho_t' = -\tau_t'\rho_t \;\; \text{(half-period thickness)} \qquad (8)$$

$$e^{-i2\xi\omega_0\Delta t} = -1 \;\Rightarrow\; \tau_t\rho_t' = \tau_t'\rho_t \;\; \text{(quarter-period thickness)} \qquad (9)$$

The first case (8) requires that $2\xi\omega_0\Delta t = 2m\pi$, where $m$ is an integer. This gives the *half-period* thickness condition $\xi\Delta t = mT/2$. In addition, using (2) and (4), the condition $\tau_t\rho_t' = -\tau_t'\rho_t$ requires that $n_0 = n_2$, that is, the media before and after the application time $\xi\Delta t$ of the temporal slab must be the same. In the special case $m = 0$, the temporal slab would have zero temporal thickness and the same refractive index before and after, which results in no medium discontinuity.

The second case requires that $2\xi\omega_0\Delta t = (2m+1)\pi$, which results into the *quarter-period* thickness condition $\xi\Delta t = (2m+1)T/4$. Furthermore, the condition $\tau_t\rho_t' = \tau_t'\rho_t$ requires $n_1^2 = n_0 n_2$.

To summarize, a temporal reflection-less slab, $\Gamma_t = 0$, can be realized through time-varying metamaterials, when the application time and the refractive indices are properly designed. It is worth to highlight that, despite both solutions vanish the reflection due to the changes of the refractive index of the medium, they exhibit the fundamental difference related to the output frequency due to the different final medium $n_2$. Indeed, the quarter-period slab is a temporal impedance matching between two different media with refractive indices $n_0$ and $n_2$, which support the propagation at frequency $\omega_0$ and $\omega_2 = n_0\omega_0/n_2$, respectively. On the contrary, the half-period slab acts as a temporal Fabry-Perot resonator within two media with the same refractive indices, *i.e.*, $n_0 = n_2$, and, therefore, no frequency shift is expected between the incident and reflected/refracted waves.

Finally, as for the zero transmission condition, the zeroes of (6) can be evaluated following a similar procedure. They theoretically correspond to a fully reflective interface that does not allow the transmission of light in forward direction at the incident operation frequency $\omega_0$. Similar to the reflection-less analysis, a zero of (6) will occur for the following conditions:

$$e^{-i2\xi\omega_0\Delta t} = +1 \;\Rightarrow\; \tau_t\tau_t' = -\rho_t\rho_t' \quad \text{(half-period thickness)} \qquad \textbf{(10)}$$

$$e^{-i2\xi\omega_0\Delta t} = -1 \;\Rightarrow\; \tau_t\tau_t' = \rho_t\rho_t' \quad \text{(quarter-period thickness)} \qquad \textbf{(11)}$$

Again (10)-(11) correspond to the *half-period* and *quarter-period* temporal slabs, respectively. Using (2) and (4) in (10), we find that in the case of half-period temporal slabs the condition $\tau_t\tau_t' = -\rho_t\rho_t'$ is never satisfied, forcing transmission to be always present. On the contrary, in the second case (11), the condition $\tau_t\tau_t' = \rho_t\rho_t'$ requires $n_1^2 = -n_0 n_2$, *i.e.*, a negative refractive index for the quarter-period slab.

*Numerical results* – The verification of the theoretical analysis and the corresponding special cases has been performed though a proper set of numerical experiments based on Finite Difference Time Domain (FDTD) technique. The temporal permittivity profile has been designed as a step-like function in time-domain at the two interfaces of the temporal slab. It is important to highlight that the shift of the temporal frequency due to the sudden change of the medium properties may lead to a strong instability of the simulations. Therefore, to ensure the convergence, the domain has to be first temporally discretized for having a sufficient number of time-steps for representing the highest frequency in the domain; and, then, the spatial discretization can be performed accordingly to the Courant stability criterion. This simple trick is to be used in any simulation where the medium properties changes with time.

As for the numerical verification, we have considered the case of half-period and quarter-period temporal slabs. First, the transmission and reflection through a half-period temporal slab, i.e. a temporal Fabry-Perot resonator, have been verified [22]. As shown in Fig. 2a, we considered a temporal slab with arbitrary refractive index $n_1$ between two media with the same refractive indices, *i.e.* $n_0 = n_2$. We have selected three different refractive indices for the temporal slab $n_1 = \{2,3,5\}$.

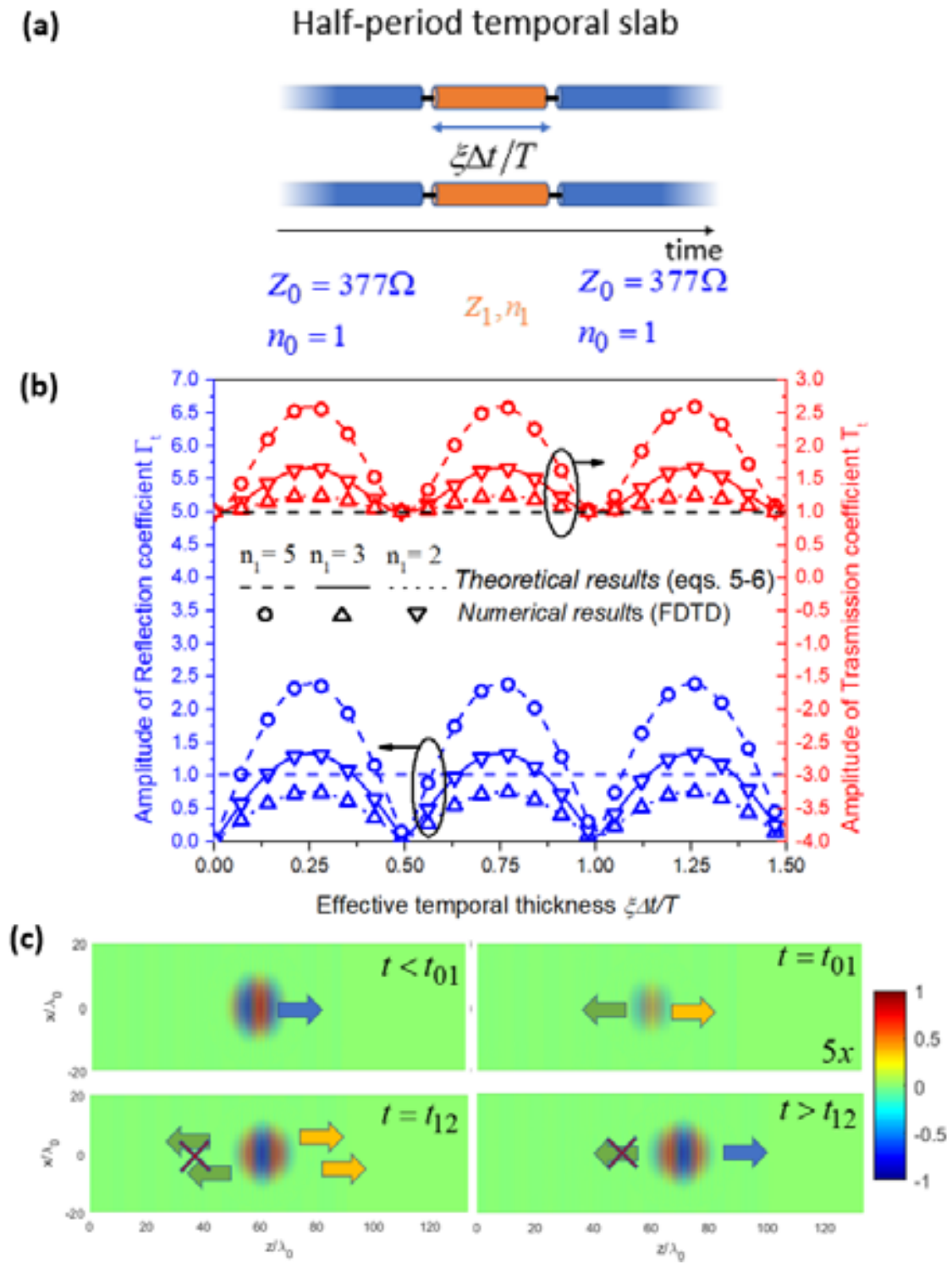

Fig. 2. (a) Equivalent temporal transmission line of the simulated 2D domain in the FDTD simulation. (b) Magnitude of the reflection and transmission coefficients as a function of the effective temporal thickness of the temporal slab with $n_1 = \{5,3,2\}$ in the case $n_0 = n_2$. (c) Snapshots in time of the 2D-FDTD simulation before (top-left), at (top-right and bottom-left), and after (bottom-right) a half-period temporal slab with $n_1 = 3$.

For each of them, the magnitudes of the transmission and reflection have been evaluated as a function of the normalized slab effective temporal thickness $\xi\Delta t / T$. As shown in Fig. 2b, and as expected from the previous theoretical analysis, the reflection coefficient vanishes periodically every half-period thickness, and, for these specific thicknesses, the transmission is unitary. Please note that for a different temporal thickness, the transmission and reflection coefficients can be larger than unity. This may be possible when temporal slabs are involved because energy conservation imposes an energy transfer from light to the medium, and vice versa, due to the frequency change [6,7]. Finally, in Fig. 2c, we show four snapshots in time before, at, and after the interfaces of the temporal slab: form top-left to bottom-right, we show the wave propagating in the medium $n_0 = 1$, at $t = t_{01}$, the wave splits due to the presence of the temporal slab with refractive index $n_1 = 3$, the side effects of the change of the refractive index are the shift of frequency to $\omega_1 = \omega_0/3$ and a drop of the wave amplitude, since the energy density strongly depends on the value of instantaneous refractive index; at $t = t_{12}$, the wave impinges on the second temporal interface and the two contributions in forward direction constructively interfere in phase (unitary transmission), whereas the two contributions in backward direction interfere destructively (zero reflection).

In Fig. 3, we report the numerical verification of the impedance matching property of a quarter-period slab. We desire to impedance match the temporal interface between a medium with $n_0 = 1$ and one with $n_2 = 3$ at the illumination frequency $\omega_0$. According to theory, the matching temporal slab should have a refractive index $n_1^2 = n_0 n_2 = 3$ (see Fig. 3a). Fig. 3b shows that the perfect matching, i.e. zero reflection, is achieved when the temporal slab has an effective temporal thickness equal to $T/4$, $3T/4$, etc. (Fig. 3b). Again, in Fig. 3c, we report four snapshots in time at the key instants of time, i.e. before, at, and after the temporal interfaces of the slab, showing again that the backward wave completely vanishes due to the destructive interference between the two backward propagating contributions.

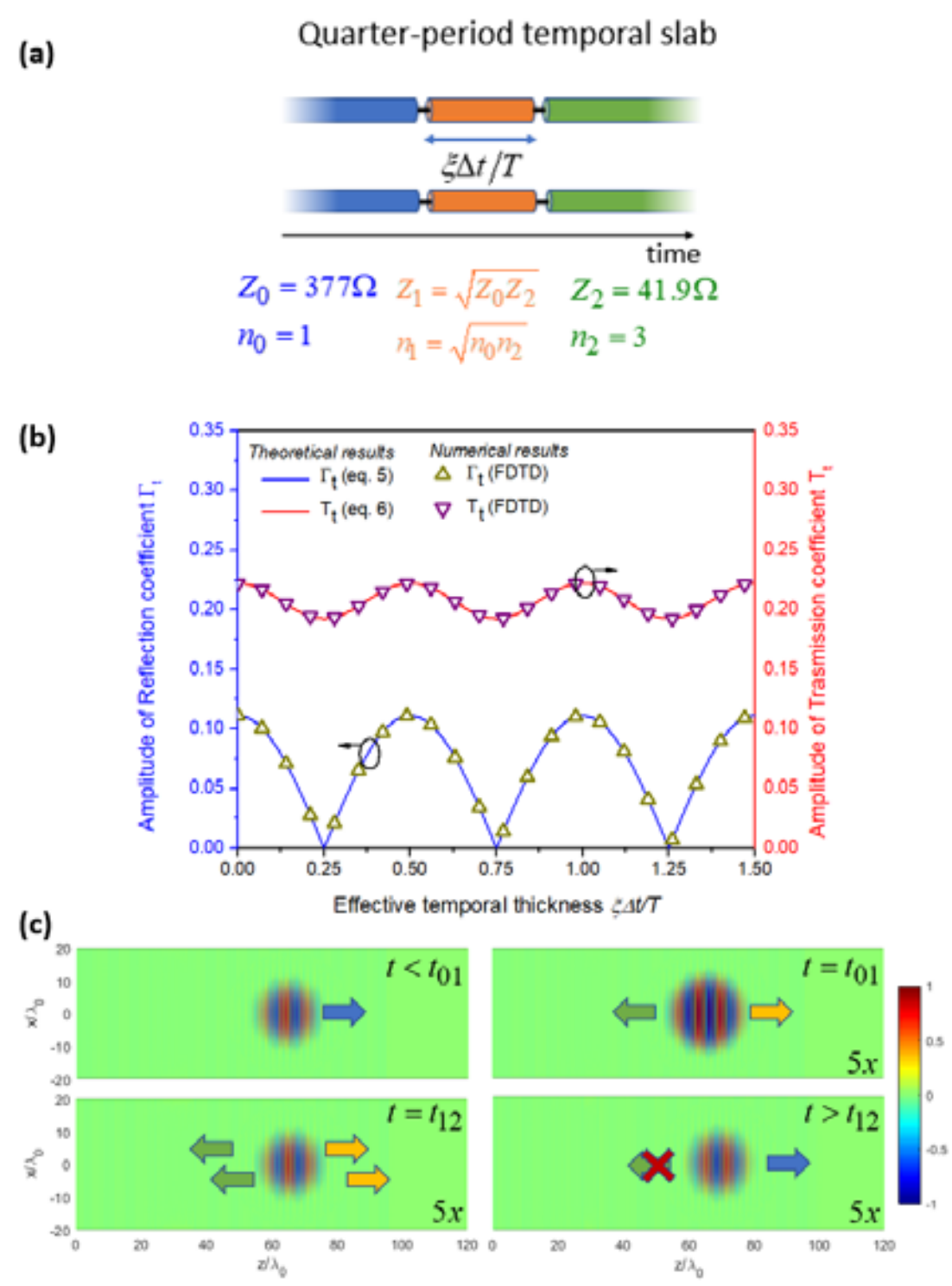

Fig. 3. (a) Equivalent temporal transmission line of the simulated 2D domain in the FDTD simulation. (b) Magnitude of the reflection and transmission coefficients as a function of the effective temporal thickness of a temporal slab with $n_1 = \sqrt{3}$ when it is between two temporal media with $n_0 = 1$, $n_2 = 3$. (c) Snapshots in time of the 2D-FDTD simulation before (top-left), at (top-right and bottom-left), and after (bottom-right) a matching quarter-period temporal slab.

*Conclusions* – To conclude, in this letter, starting from the earlier works on temporal discontinuities, we have proposed a non-trivial extension of the study to multiple temporal discontinuities, proposing the concept of *temporal slab*. We shed light on the scattering from a metamaterial temporal slab, deriving the reflection and transmission coefficients as a function of the refractive indices and the application time. We have shown that the response of the temporal slab can be controlled through the application time, which acts similarly to the electrical thickness of conventional spatial slabs. We have demonstrated that temporal Fabry-Perot resonators and temporal anti-reflection coatings can be successfully implemented. The basic temporal components identified in this letter may pave the way to the design of several novel devices based on temporal discontinuities such as temporal matching networks, Bragg grating, dielectric mirrors, which exhibit zero space occupancy by exploiting the time dimension.

**Disclosures**. The authors declare no conflicts of interest.